\documentclass[10pt,conference]{IEEEtran}
\usepackage{cite}
\usepackage{amsmath,amssymb,amsfonts}
\usepackage{algorithmic}
\usepackage{graphicx}
\usepackage{textcomp}
\usepackage{xcolor}
\usepackage{listings}

\usepackage[%
PlaceholderOption]{drafthelper}

{\begin{small}\begin{list}{0}{%
\setlength{\topsep}{0.0pt}%
\setlength{\itemsep}{0.6pt}%
\setlength{\leftmargin}{0.15in}%
\setlength{\itemindent}{0.0in}%
\setlength{\rightmargin}{0.075in}%
\setlength{\labelsep}{0.25em}}}%
{\end{list}\end{small}}

{\begin{list}{0}{\setlength{\leftmargin}{0.2in}\setlength{\itemindent}{0.0in}\setlength{\labelsep}{0.25em}\setlength{\itemsep}{0pt}}}%
{\end{list}}

{\begin{list}{$\bullet$}{\setlength{\leftmargin}{0.05in}\setlength{\labelsep}{0.5em}}}%
{\end{list}}

\def\BibTeX{{\rm B\kern-.05em{\sc i\kern-.025em b}\kern-.08em
    T\kern-.1667em\lower.7ex\hbox{E}\kern-.125emX}}
\begin{document}

\title{An Inquisitive Code Editor for Addressing Novice Programmers' Misconceptions of Program Behavior}

\author{

\IEEEauthorblockN{Austin Z. Henley, Julian Ball, Benjamin Klein, Aiden Rutter, Dylan Lee}
\IEEEauthorblockA{
University of Tennessee\\
Knoxville, Tennessee \\ 
azh@utk.edu, \{jball16, bklein3, arutter1, dlee97\}@vols.utk.edu}

}


\maketitle


\begin{abstract}
Novice programmers face numerous barriers while attempting to learn how to code that may deter them from pursuing a computer science degree or career in software development. In this work, we propose a tool concept to address the particularly challenging barrier of novice programmers holding misconceptions about how their code behaves. Specifically, the concept involves an inquisitive code editor that: (1) identifies misconceptions by periodically prompting the novice programmer with questions about their program's behavior, (2) corrects the misconceptions by generating explanations based on the program's actual behavior, and (3) prevents further misconceptions by inserting test code and utilizing other educational resources. We have implemented portions of the concept as plugins for the Atom code editor and conducted informal surveys with students and instructors. Next steps include deploying the tool prototype to students enrolled in introductory programming courses.
\end{abstract}







\section{Introduction}
\label{sec-introduction}

\Intent{The big problem: Novice programmers often hold false assumptions about how their code behaves.}
%
%
Novice programmers face numerous barriers while attempting to learn how to code that may deter them from pursuing a computer science degree or career in software development.
In this work, we propose a tool concept to address the particularly challenging barrier of novice programmers holding misconceptions about how their code behaves.
A key reason that novice programmers have this issue is they often have a different definition of program correctness than professional programmers~\cite{Kolikant2005ICER}.
In fact, one study found that many students believe a program is correct when there are no compiler errors, but the students indicated little regard to the program's behavior~\cite{Stamouli2006ICER}.
%
These \emph{semantic errors}, when the program compiles but behaves differently than the programmer expects, have vexed many students, and in another study, it took students substantially longer to fix semantic errors than syntax errors (if they were fixed at all)~\cite{Altadmri2015SIGCSE}. 
%

\Intent{Our big idea: an inquisitive code editor.}
Towards addressing these issues, we propose an inquisitive code editor that elicits information about program behavior from novice programmers to address their misconceptions.
The editor will prompt them with questions about the program's behavior that can be verified using static analysis and utilize the response as a learning opportunity for the programmer.
%
%
More specifically, the editor will do this by detecting potentially buggy code using a code smell detector that is attuned to mistakes made by novices. 
Then it will prompt the programmer with a question about the program behavior that can be validated with static analysis. 
If the programmer answers incorrectly, the tool will provide an explanation about the program behavior and insert assertions and comments into the code.
Given these features, the goals of the inquisitive code editor are to (1) identify when a programmer has a misconception about the program behavior, (2) correct the misconception by explaining how the program will actually behave, and (3) prevent the programmer from having similar misconceptions again.
%
%
%
%

The foundation of our tool concept is based on prior research on debugging tools, empirical evidence of mistakes that novice programmers make, and tutoring systems.
A notable research tool for debugging is Whyline, which enables programmers to ask \emph{why} and \emph{why not} questions about their program's behavior~\cite{Ko2004CHI, Ko2008ICSE}.
However, this may not be suitable for novices since Whyline requires that the programmer already knows that a bug exists and is able to replicate the buggy behavior.
The aim of our inquisitive code editor is to instead have the code editor asking the questions, such that novice programmers have a more efficient feedback loop regarding their program's behavior and can effectively address their misconceptions.
Moreover, our concept makes use of recent research on tutoring systems for learning to code (e.g., PLTutor~\cite{Nelson2017ICER} and Ludwig~\cite{Shaffer2005SIGCSE}).

\Intent{In this paper.}
In this paper, we present a detailed description of our novel tool concept along with a mock implementation in the form of a plugin for the popular Atom code editor (Section II).
We also share our preliminary results and efforts towards researching the effectiveness of our technique (Section III).
Following that, we provide a detailed literature review of relevant background information and related works (Section IV).
Finally, we conclude with the next steps in our research towards investigating how to better address the misconceptions of novice programmers (Section V).

\DraftPageBreak{}


\section{Idea: An Inquisitive Code Editor}
\label{sec-idea}

\Intent{Our big idea.}
To address the misconceptions that novice programmers have about their code, we propose a novel concept of an inquisitive code editor that elicits information about program behavior from novice programmers to address their misconceptions.
The editor concept does this through a three step process.
First, it can identify misconceptions by periodically prompting the programmer with questions about the program's behavior.
Second, it can correct the misconceptions by generating explanations based on the program's actual behavior.
Third, it can prevent further misconceptions by inserting test code (e.g., assertions or unit tests) and documentation, providing external help for more in-depth explanations, and send aggregate reports to the instructor.
%
%
%

\begin{figure}
\centering
\includegraphics[scale=0.95]{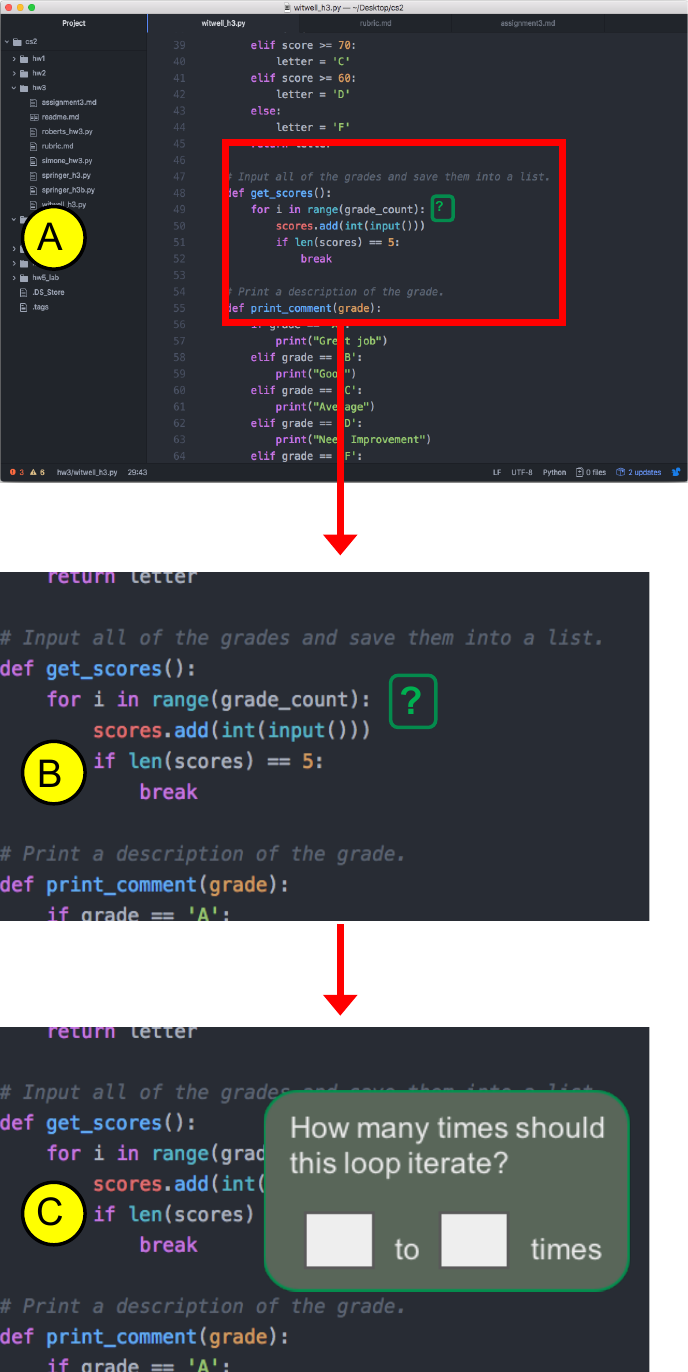}

\vspace{-7pt}
\caption{%
%
(A) A mockup of the inquistive code editor for addressing misconceptions as a plugin for the Atom code editor.
(B) Code that may be misunderstood is automatically annotated with a green question mark.
(C) Clicking the annotation displays a question for the programmer to answer about how they think the program will behave. 
}
\label{fig-tool}
\end{figure}

\subsection{Identifying Misconceptions}

\Intent{}
The technique that we propose is for the editor to \emph{ask} the programmer questions about the behavior of specific portions of code while the programmer is writing the code.
For example, it can ask, ``How many times will this loop iterate?'', along with a textbox or slider for the programmer to input the range they believe is correct.
Using \emph{program analysis}, the tool can compare the programmer's answer to how the analysis determined that the program will actually behave.
This example question could identify common loop misconceptions, such as a loop that will only ever iterate one time.
\CalloutFigure{fig-tool}-A depicts a mockup of a tool that extends a code editor to annotate potentially problematic code (\CalloutFigure{fig-tool}-B) and pose questions to the programmer about the code (\CalloutFigure{fig-tool}-C).

\Intent{}
There are a number of design dimensions that need to be studied in order to identify misconceptions effectively.
First, which misconceptions and code issues should the code editor ask questions about (and how does it know this is a potential bug)?
There are existing databases of bugs that students have encountered in various course settings, such as Blackbox~\cite{Brown2014SIGCSE}.
The tool could combine such a database with existing code smell detector techniques.
Second, how does the tool determine the correct answer to how the program will behave?
There are a variety of program analysis techniques that could be used (e.g., data-flow analysis) based on the misconceptions targeted.
These static and dynamic program analysis techniques allow the editor to infer specific properties about the program behavior (e.g., a loop will never terminate).
Third, how should programmers interact with the editor without finding it distracting?
From our prior work, we have found that it is vital for the tool to work without upfront configuration and without disruption to their workflow.
For example, Yestercode and CodeDeviant record code changes automatically and provide visualizations that are available if the programmer needs assistance, without upfront configuration~\cite{Henley2016VLHCC, Henley2018VLHCC}.
If our inquisitive editor is too disruptive, the programmer will be less likely to use it and it may interfere with their task, and thus contradict the goals we are trying to achieve.
Fourth, what other techniques could be used in conjunction with program analysis for identifying misconceptions?
Other researchers have began looking into ways to identify when a programmer is frustrated (e.g., ~\cite{Drosos2017VLHCC, Muller2015ICSE, Rodrigo2009ICER}), which could also be indicative of a misconception.
Additionally, other signals could be investigated to see if they correlate with misconceptions (e.g., keyboard and mouse activity, inserting breakpoints, or running the program repeatedly).

\Intent{}
As such, our initial design is to annotate the potentially problematic code in the code editor (shown in \CalloutFigure{fig-tool}-B as a green question mark), much like a syntax error.
Annotating the code is a deliberate alternative to interrupting the programmer with a dialog window whenever the issue is identified.
These annotations are based on \emph{negotiable interruptions}, notifications that allow the user to attend to when they choose, and studies have found them to be preferred in long, continuous tasks~\cite{McFarlane2002HCI}.
\CalloutFigure{fig-tool}-C shows an example of a question that is posed to the programmer after clicking the green question mark, indicating a portion of code that may be indicative of misconceptions.
There are a variety of different question types and input forms (e.g., fill in the blank, drag-and-drop, natural language, etc.) that should be investigated.

\subsection{Correcting Misconceptions}

\Intent{}
After identifying a misconception, our inquisitive code editor aims to then \emph{correct} programmers' misconceptions.
To do so, the editor will present an explanation as to how the program will actually behave and why the program will behave that way.
Using the programmer's answer to the question (e.g., \CalloutFigure{fig-tool}-C), the editor will generate a customized explanation based on the specific misconception.
More specific designs can utilize findings from \emph{computing education research}, such as tutoring systems, hint generation, and enhanced compiler errors.

\Intent{}
There are two open questions regarding the features for correcting programmers' misconceptions: \emph{what} information to provide and \emph{how} to present it.
It is likely not sufficient to just inform the programmer how the program will behave (e.g., ``This loop will never iterate more than 5 times.'').
An existing tool, the Whyline, was designed to help programmers interrogate their program to understand why it is not behaving as expected~\cite{Ko2004CHI}.
However, our goal is different in that we want to assist the novice programmer in overcoming a misconception that they do not even know they have yet.
The information needs of novice programmers is likely dependent on the situation and context, and may need to be adapted based on the environment.

\Intent{}
Furthermore, an interactive explanation may lead to more engagement and better understanding for the programmer.
For example, the editor could tell the programmer to test the program with specific inputs that would cause it to fail, and thus, give the programmer a first-hand experience of the program not behaving as they expected.
A similar approach, known as \emph{surprise-explain-reward}, has been applied to programming environments with notable benefits~\cite{Cao2015IWC, Jernigan2015VLHCC}.
For example, it has been used to entice end-user programmers in testing spreadsheets by inputting arbitrary values into the spreadsheet to see if it ``surprises'' the programmer with the result~\cite{Wilson2003CHI}.
Seeing the program behave in an unexpected manner could have a positive effect on the programmer's understanding rather than just reading an explanation, and also encourage the programmer to continue testing the program.
This approach should prove to be an improvement over just providing textual explanations, such as enhanced compiler error messages, which have yielded mixed results~\cite{Becker2016SIGCSE, Denny2014ITiCSE, Flowers2004FIE, Nienaltowski2008SIGCSE, Pettit2017SIGCSE, Prather2017ICER}.

\subsection{Preventing Misconceptions}

\Intent{}
Once the programmer's misconception has been identified and corrected, our proposed code editor will attempt to prevent the programmer from having similar misconceptions again.
There are three mechanisms that can be utilized to prevent programmers from having similar misconceptions in the future about their program's behavior.
First, the editor could generate test code (i.e., assertions or unit tests) and documentation (i.e., code comments).
The goal of the test code is to inform the programmer if and when the program behavior changes, such that their understanding of the program's behavior changes accordingly.
For example, an assertion could be inserted that will cause an immediate error if a variable does not always fall within a specified range.
For more advanced students, the tools could automatically generate unit tests, a more sophisticated means of testing than assertions, but are often not taught in introductory computer science courses.
The tools will also automatically generate code comments to provide the programmer a reminder of how the code behaves and an explanation as to why it behaves that way.
To avoid causing additional frustration and cognitive load, the tools will need to provide features to efficiently hide the inserted test code and comments, such as a toggle button.
A concern is that inserting too many lines of code or comments could make navigating the code tedious and difficult, so this will require further investigation on how to do so efficiently and succinctly.

\Intent{}
Second, the editor could monitor which types of issues the programmer is struggling with (e.g., loops or bitwise operators) and use it as an opportunity to teach that topic to the programmer.
For example, if multiple misconceptions have been identified regarding the number of iterations in a loop, the tool could walk through each piece of a loop with explanations on the syntax and semantics.
If more help is needed, the tools could provide excerpts from or links to relevant code documentation, programming tutorials, and examples.

\Intent{}
Third, the code editor could provide aggregate feedback to course instructors on what misconceptions their students are often facing.
To do so, the editor could be configured to anonymously collect data on misconceptions in a course setting (e.g., homeworks or in-class activities) and upload the data to a web server that is accessible by the course instructor.
The course instructor could then view reports of what programming constructs are students having difficulty in understanding and adapt the course to provide additional instruction on the problematic topics.

\DraftPageBreak{}


\section{Preliminary Results}
\label{sec-results}

\Intent{}
We have been investigating our tool concept from a number of different perspectives.
We have implemented a plugin for Atom that supports annotating potentially problematic code snippets, presenting a question and allowing the programmer to enter an answer, and presenting an explanation and inserting an optional code fix or assertion.
Additionally, we have created a framework for logging interaction data in the code editor that is submitted to a remote server in a confidential manner, such that we can later analyze the effectiveness of our technique.
%

To identify the potentially problematic code snippets, we first investigated the viability of off-the-shelf code smell detectors and linters (e.g., SonarSource\footnote{https://rules.sonarsource.com/}).
However, these tools are predominately designed for professional software developers.
By examining the rules employed by such tools and showing them to undergraduate students, we found that the majority of them are of little to no value to novice programmers and could actually cause further confusion.
Furthermore, these tools do not report many of the semantic errors that seem prevalent among novices.

We are exploring the design of code smell detectors that are specific to mistakes that novices make.
While it is trivial to detect individual mistakes reported by instructors and researchers (e.g., \cite{Altadmri2015SIGCSE}), this may not generalize to other programming languages or programming assignments.
In an effort to overcome this limitation, we have combined several large data sets, including Blackbox~\cite{Brown2014SIGCSE}, Stack Overflow, and GitHub.
Using this data, we are currently training various classifiers to detect patterns in code that has indicators of being buggy or not buggy.

Our informal surveys on the usefulness of our inquisitive code editor have provided us positive feedback.
A primary concern is that students would ignore or be annoyed by the tool's questions, but due to the different workflow and mindset of novices, they seem to be eager for any opportunity to receive help with their code.
Additionally, we asked how students felt about a tool that would insert code comments or assertions into their code as they worked, and they responded favorably.
When speaking with instructors for introductory programming courses, they believe such a tool could greatly benefit students, so long as the tool does not solve the bugs for them or let the students become dependent on it.

\DraftPageBreak{}



\section{Related Work}
\label{sec-relatedwork}

\subsection{Mistakes Novice Programmers Make}

\Intent{Novices make a lot of syntax mistakes, and here is what they are like.}
Novice programmers make a lot of mistakes in their code~\cite{Altadmri2015SIGCSE, Becker2016ITiCSE, Brown2014ICER, Ebrahimi1994IJHCS, Hristova2003SIGCSE, Jackson2005FIE, Spohrer1986CACM}.
Although the majority of the mistakes are syntax errors (e.g., unbalanced parentheses), these errors are often fixed in a relatively short time~\cite{Altadmri2015SIGCSE}.
One common approach to aiding students with syntax errors is for tools to provide \emph{enhanced compiler error messages} that are easy to comprehend and may provide suggestions as to how to address the error.
However, studies so far have had inconsistent or inconclusive results as to how effective they are in benefiting students~\cite{Becker2016SIGCSE, Denny2014ITiCSE, Flowers2004FIE, Nienaltowski2008SIGCSE, Pettit2017SIGCSE, Prather2017ICER}
Another approach that is growing in popularity is blocks-based languages (e.g., Scratch~\cite{Maloney2010TOCE} and App Inventor~\cite{Wolber2011Book}), which mitigate syntax errors completely by using a drag-and-drop interface that does not require memorizing syntax, but these languages have not yet been adopted in industry.

\Intent{}
Semantic errors---when the program behaves differently than expected---are also common~\cite{Altadmri2015SIGCSE, Brown2014ICER} and even more challenging to fix since compilers often do not provide any warnings about them.
In fact, one study found that students took far longer to fix semantic errors than syntax errors, if they were fixed at all~\cite{Altadmri2015SIGCSE}.
The underlying cause of why novices introduce semantic errors is complex, but may stem from a lack of conceptual knowledge, inadequate problem solving strategies, or extraneous cognitive load~\cite{Qian2017TOCE}.
Consider the following snippet of Python code based on a student's homework submission:
\lstset{}
\begin{lstlisting}[basicstyle=\small,language=Python, columns=fullflexible]
response = input(``Please enter (y)es or (n)o'')
while response != 'y' or response != 'n':
    response = input(``Please enter (y)es or (n)o'')
\end{lstlisting}
From the student's perspective, this code should work.
The program will ask the user to input `y` or `n`, and if anything else is typed in, the program will continue to ask for the correct input.
However, this contains a semantic error since the logical ``or'' in the loop condition will cause the loop to iterate forever.

\Intent{}
Researchers have proposed other techniques and tools that may alleviate various errors that novices make.
One such approach is to provide visualizations of how the program will behave to aid programmers in understanding.
A notable example of such a tool is Python Tutor, which displays visualizations of the program's data structures at each step of the code~\cite{Guo2013SIGCSE}. 
Other tools include Omnicode, which displays a scatterplot matrix of all run-time values for every variable in the program~\cite{Kang2017UIST}, Theseus, which annotates functions in the code editor with the number of times it was called during the current execution~\cite{Lieber2014CHI}, and a tool extension that displays a small graph of how each variable changes over time during execution~\cite{Hoffswell2018CHI}.
Another approach is to generate content to help programmers, such as hints~\cite{Galenson2014ICSE, Head2015VLHCC, Ichinco2018IDC, Peddycord2014EDM}, examples~\cite{Ichinco2016VLHCC, Ichinco2017CHI, Oney2012CHI}, tutorials~\cite{Harms2013IDC}, and recommendations~\cite{Fast2014CHI, Hartmann2010CHI, Mujumdar2011CHI, Suzuki2017VLHCC, Watson2012ICWBL}.
Recently, researchers have began exploring how to enable users to efficiently tutor students over the web~\cite{Guo2015UIST, Guo2015VLHCC, Warner2017CHI}.
These techniques could all benefit our proposed work in overcoming misconceptions.

\subsection{Asking Questions about Program Behavior}

\Intent{Whyline}
A particularly relevant tool to our proposed research is the Whyline~\cite{Ko2004CHI, Ko2008ICSE, Ko2009CHI}, which enables programmers to record program executions and then ask ``why'' and ``why not'' questions about the program's behavior.
For example, a programmer can use the tool to click on a specific part of their program's interface and ask the question, ``why is this button red?''
The Whyline will then provide an answer (i.e., the relevant code) and allow the programmer to ask follow-up questions, such as ``why did this event not occur?''
To provide these answers, Whyline records an extensive amount of information from a program's execution and uses a combination of static and dynamic program slicing to derive the answers about the program behavior~\cite{Ko2008ICSE}.
%
%
%
In an empirical study, participants using Whyline were able to fix bugs in half of the time than those using a traditional debugger~\cite{Ko2009CHI}.

\Intent{However, Whyline...}
Despite the great benefits that the Whyline provides, there a number of ways in which it does not solve the misconceptions that we are striving to overcome for novice programmers.
One potential barrier is that using the Whyline requires the programmer to have already acknowledged that there is a bug and to have recorded the executing program while displaying the buggy behavior.
This is insufficient since novices may never even realize that there is a bug in their code.
Additionally, novices may find it difficult to find a location in the program that is relevant, then form a question to ask.
Beyond preemptively catching misconceptions, our work also aims to correct and prevent misconceptions.
Although the Whyline enables the programmer to trace the lines of code that may be relevant to a bug, it does not provide any additional explanation that may be needed for a novice to understand. 
Moreover, the Whyline does not provide any features that prevent the programmer from having similar misconceptions in the future.
Even though the Whyline has made tremendous progress in improving the debugging process, it is clear that there is still a gap in tools that support novices in overcoming semantic errors.

\subsection{Tutoring Systems}

\Intent{}
To provide explanations and feedback about the program behavior, we will utilize findings from research on \emph{tutoring systems}.
These systems provide some form of instruction or feedback to students as if it was an automated teacher.
One such system, PLTutor, provides short lessons along with a code editor and visualization of the program state~\cite{Nelson2017ICER}.
Another tutoring system is Ludwig~\cite{Shaffer2005SIGCSE}, which provides students feedback on code style and compares the student's program output to the instructor's program output.
Furthermore, a number of programming tools have augmented features of tutoring systems into code editors, such as providing hints to novices in open-ended programming tasks~\cite{Ichinco2018IDC}.
AutoTutor, a well-studied tutoring system for many subjects other than programming, has shown great promise in keeping students engaged and identifying students' emotions using mixed-initiative dialog in natural language~\cite{DMello2013TiiS, Graesser2005TOE}. 
These techniques could be essential for the tool to be used effectively by novice programmers.


%

\DraftPageBreak{}


\section{Conclusion}
\label{sec-conclusion}

\Intent{}
The next major steps in our research towards making this idea a reality include conducting a laboratory study and a field study.
With a lab study involving our inquisitive code editor, we can understand how users interact with the questions and the effectiveness in how it corrects misconceptions in a controlled environment.
To better understand the usefulness of our tool in a more ecologically valid setting, we will deploy it to students enrolled in introductory programming courses to be used while completing their homework assignments.
The logging framework that we developed will provide us data on what bugs the students have, the changes they make to fix the bug, how they debug, and how they interact with our tool.

\Intent{Sunset. We shall save the world!}
By proposing this initial tool concept and conducting the research necessary to implement the inquisitive code editor, we strive to advance knowledge in the areas of \emph{software engineering}, \emph{human-computer interaction}, and \emph{computing education} in order to help novice programmers understand what their code is really doing.
With the ever growing need for more people with computer programming skills, it is of paramount concern to reduce the barriers that may deter people from obtaining these skills.
Understanding code is an already arduous process that students go through, thus our work aims to alleviate these difficulties and to support a diverse population of students in learning how to program.

\DraftPageBreak{}


\section{Acknowledgments\DraftStatus{0}\PageBudget{0}}

This material is based upon work supported by the National Science Foundation under grants 1850027 and 2008408.




\bibliographystyle{IEEEtranS}
\bibliography{refs}
%



\end{document}